\documentclass[12pt]{spieman}  
\usepackage{amsmath,amsfonts,amssymb}
\usepackage{graphicx}
\usepackage{float}
\usepackage{setspace}
\usepackage{tocloft}

\title{Decentralized Smart Surveillance through Microservices Platform}

\author[a]{Seyed Yahya Nikouei}
\author[a]{Ronghua Xu}
\author[a,*]{Yu Chen}
\author[b]{Alex Aved}
\author[b]{Erik Blasch}
\affil[a]{Dept. of Electrical \& Computer Engineering, Binghamton University, Binghamton, NY 13902}
\affil[b]{The United State Air Force Research Laboratory (AFRL), Rome, NY 13441, USA, 13441}

\cftpagenumbersoff{figure}
\cftpagenumbersoff{table} 

\begin{document} 
\maketitle

\begin{abstract}
Connected societies require reliable measures to assure the safety, privacy, and security of members. Public safety technology has made fundamental improvements since the first generation of surveillance cameras were introduced, which aims to reduce the role of observer agents so that no abnormality goes unnoticed. While the edge computing paradigm promises solutions to address the shortcomings of cloud computing, e.g., the extra communication delay and network security issues, it also introduces new challenges. One of the main concerns is the limited computing power at the edge to meet the on-site dynamic data processing. In this paper, a Lightweight IoT (Internet of Things) based Smart Public Safety (LISPS) framework is proposed on top of microservices architecture. As a computing hierarchy at the edge, the LISPS system possesses high flexibility in the design process, loose coupling to add new services or update existing functions without interrupting the normal operations, and efficient power balancing. A real-world public safety monitoring scenario is selected to verify the effectiveness of LISPS, which detects, tracks human objects and identify suspicious activities. The experimental results demonstrate the feasibility of the approach.
\end{abstract}

\keywords{Smart Surveillance, Microservices Architecture, Decentralization}

{\noindent \footnotesize\textbf{*}Corresponding Author: Yu Chen,  \linkable{ychen@binghamton.edu} }





\section{Introduction}
\label{sect:intro}  

Modern cities are more congested than ever because of citizen life style choices and urban opportunities \cite{zlotnik2017world}. The fast speed of urbanization also brings concerns for public safety as there infrastructure, people, and network designs \cite{chen2016smart}. Nowadays, a lot of video cameras are deployed in cities to monitor people's activity and movements. Normally, surveillance cameras are connected to the network to deliver the video to a central location or a server for further analysis. 
Although the internet infrastructure is getting more robust, the workload is getting heavier and heavier. Online video accounted for 74\% of all online traffic in 2017 \cite{index2016forecast} and 78\% of mobile traffic will be video data by 2021 \cite{meeker2017internet}. Obviously, if the video can be processed on-site or near-site at the edge, close to where the video sensors are installed, unnecessary delays and overhead on communication network will be significantly reduced.

Being aware of the growing demand for resources due to the ubiquitous deployment of networked static and mobile cameras, the surveillance community has made many efforts in past decades~\cite{ma2017survey}. The cloud, fog and edge computing paradigms are adopted in new surveillance systems. In the \emph{cloud computing} based systems the cameras are connected to a cloud center that may be located geographically miles away. The video footage is streamed to the cloud where it will be stored and processed for future analysis. 
Huge investment is made every year to keep the data centers and storage up and running. Also, because of the overwhelmingly large number of cameras, as the video is captured, it is unlikely that an administrator is able to timely narrow down to the frames according to the scenarios of an emergency. In addition, there are challenges to assure the security of the data. In practice, the video is often sent through the network without sufficient security measures. The communication can be intercepted and a third party can steal the video or even worse to launch frame duplication attacks \cite{nagouthu2019detecting, nagouthu2019study}. 

\begin{figure}[t]
    \centering
        \includegraphics[width=0.5\textwidth]{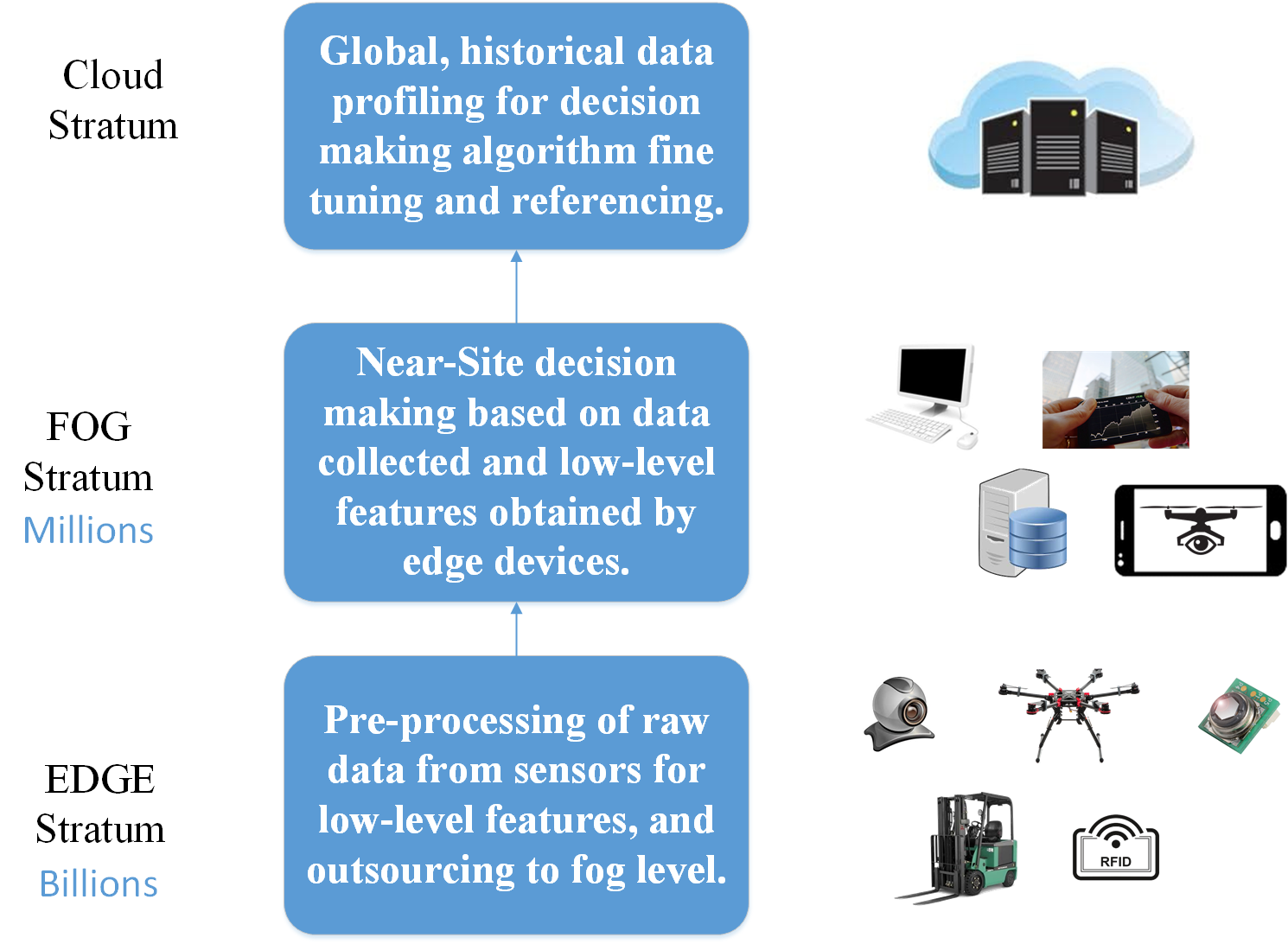}
    \caption{The Edge-Fog-Cloud hierarchical architecture.}
    \label{fig:edge-fog-cloud}
\end{figure}

Migrating more computing tasks to the connected smart “things” (sensors and actuators) at the edge of the network \cite{shi2016edge}, \emph{edge computing} is considered as the answer to the shortcomings of cloud computing \cite{ahmed2017mobile, cao2017edge}. The process takes place on-site or near-site at the location of the sensor rather than uploading sensing data to the cloud \cite{chen2017enabling}. The hierarchy is illustrated in Fig. \ref{fig:edge-fog-cloud}. The first level of the processing is applied close to the sensor and more complex computation is offloaded to a fog node, and the cloud node is used for long term analysis and forensics. 

One challenge of edge computing paradigm lies in the performance gap between the limited computing power at the edge and the requirements for instant dynamic data processing. \emph{Fog computing} provides the balance between the cloud computing and edge computing paradigms. Fog computing uses edge devices that transmit data through a local area network (LAN) gateway that provides the processing and analysis services. The fog server collects data from the edge devices and services which subsequent analysis is sent to a cloud center. 

In this paper, a Lightweight IoT based Smart Public Safety (LISPS) framework is proposed using a microservices architecture to support complex public safety tasks at the edge and fog levels. Leveraging the advanced features of microservices architecture, complex system functions are decoupled into smaller and more manageable sub-tasks. Therefore, the LISPS system possesses high flexibility both in the design process and online maintenance. The loose coupling architecture enables developers to add new services or update existing functions without interrupting normal operation. 


The rest of the paper is as follows. Section \ref{sect:back} presents the background knowledge and related work. Section \ref{sect:overview} discusses the high level LISPS architecture of the framework and in Section \ref{sec:decison} the details of each part are described. Section \ref{sec:exp} will present our findings in some real life scenarios. Lastly, Section \ref{sec:conclusions} gives final remarks and discussions of the on-going efforts.

\section{Related Work}
\label{sect:back}

This section first introduces the microservices architecture and the background knowledge of Docker, where Docker is one of the most popular platforms for microservices development. Further clarification for surveillance applications at the edge is discussed followed by the security and distribution processing of edge devices.

\subsection{Microservices Architecture}


Microservices architectures (MSA) are extensions of a Service Oriented Architecture (SOA) \cite{wilde2016approaches}. The traditional SOA uses a monolithic architecture that constitutes different software features in a single interconnected application and database. Relaying on the tightly coupled dependence among functions and components, it is more challenging to adapt to new requirements in an IoT-enabled system than the SOA, which requires scalability, service extensibility, and cross-platform interoperability \cite{datta2018next}. The main difference between MSA and SOA is that a MSA is more decentralized and it distributes all the logic (such as routing, message parsing) in to smart end points \footnote{http://www.opengroup.org/soa/source-book/msawp/p3.htm}. MSA also adopts a light weight applications programming interface (API) gateway for managing services instead of using heavier and more sophisticated Enterprise Service Bus \cite{lander2019microservices}. Each microservice runs its own process and communicates with peers using light weight communication mechanisms (REST API or SOAP protocol \cite{malavalli2015scalable}, etc). The services are built around business capabilities and independently deployed by fully automated deployment tools \cite{pathania2017setting}.

Fine granularity and loose coupling are two significant features in MSA \cite{yu2018survey}. Fine granularity means that there is a bare minimum of centralized management of these services. Moreover, there are instances where the service is not micro \cite{sheng2019micro}. In the smart safety system, some parts of the code is inseparable which makes a bigger service. Microservices will become as big as it needs to be to provide coherent, reliable, and efficient functionality. Loose coupling implies that each of microservices components has few dependencies on other separate components, which makes the deployment and development of Micro-services more independent \cite{pautasso2017microservices}.

Thanks to multiple attractive properties for distributed computing, such as self-containment, coupling, orchestration, and extensibility, the MSA has been utilized in many scenarios to enhance the scalability and security of IoT-based applications \cite{datta2018next}, including Intelligent Transportation Systems (ITS) \cite{herrera2018smart}, smart city IoT platform \cite{krylovskiy2015designing}, smart building \cite{bao2016microservice}, etc. To enable more decentralized and scalable solutions for advanced video stream processing for large volumes of distributed edge devices, a conceptual design of a robust smart safety system was proposed earlier \cite{nikoueiposter}. Based on the MSA that incorporates blockchain technology for added security \cite{nagothu2018microservice}, the LISPS aims at offering a scalable, decentralized and fine-grained smart surveillance systems. This paper advances the LISPS methods for a smart public safety (SPS) platform, where extensive results validate the system performance and efficiency over conventional SOA methods.


\subsection{Smart Surveillance}

There is a growing demand for human resources to interpret the live data streams from security cameras \cite{blasch2012high}. Numerous automated object detection algorithms have been investigated to atomize this process using statistical analysis~\cite{fuse2017statistical} or machine learning (ML)~\cite{ribeiro2017study} approaches. The ML methods are computationally expensive such as algorithms normally implemented at the powerful cloud servers of the surveillance system. For example, Wide Area Motion Imagery (WAMI) transforms imagery from sensors back to the cloud for processing, which puts a heavy burden on the communication network and all aforementioned problems \cite{wu2017container, wu2015pseudo}. 
To reduce data transmission, it has been suggested to promote operators' awareness by using context \cite{blasch2014context} or providing query languages \cite{aved2015multi}. Further, approaches such as re-configuring the networked cameras \cite{piciarelli2016dynamic}, utilizing event-driven visualization \cite{fan2017heterogeneous}, and mapping conventional real-time images to 3D camera images \cite{jiewu1} improve the efficiency and throughput of the communication networks along better detection rates. 

Decentralized surveillance systems are more suitable in many mission-critical, delay sensitive tasks \cite{nikouei2018eiqis}. Recent developments of the edge hierarchy architecture enables real-time surveillance based on the fog computing paradigm \cite{mahmud2018fog}. Many on-line and uninterrupted target tracking systems are proposed to meet the requirements of real-time video processing and instant decision making deployed at the edge \cite{mukherjee2018survey}. Researchers also merged raw video streams from drones on near-site fog computing devices to reduce the amount of data to be outsourced to the cloud center \cite{chen2016dynamic}.

A safety system focusing on object assessment can be constructed following the edge-fog-cloud hierarchy \cite{mouradian2017comprehensive}. The input surveillance camera frame is given to an edge unit where low-level processing is performed, such as feature detection and object tracking \cite{blasch1998simultaneous, howard2017mobilenets, nikouei2018intelligent}. The intermediate-level is in charge of action recognition, behavior understanding, and decision making like abnormal event detection, which is implemented at the fog stratum \cite{nikouei2018kerman}. Finally, the high-level is focused on historical pattern analysis, algorithm fine tuning, and global statistical analysis, which depends on the type of decisions the system is going to make. 

\subsection{Blockchain-enabled Security}

As a fundamental technology of Bitcoin \cite{nakamoto2008bitcoin}, blockchain initially was used for new cryptocurrencies that perform commercial transactions among independent entities without relying on a centralized authority. Essentially, the blockchain is a public ledger based on consensus rules to provide a verifiable, append-only chained data structure of transactions. Due to the decentralized architecture, blockchain allows the data to be stored and updated distributively, which makes blockchain an ideal architecture to ensure distributed transactions among all participants in a trustless environment, like edge-based IoT networks.

Emerging from the intelligent property, a \textit{smart contract} allows users to achieve agreements among parties through a blockchain network. By using cryptographic and security mechanisms, smart contract combines protocols with user interfaces to formalize and secure relationships over computer networks \cite{szabo1997formalizing}. Through exposing public functions or application binary interfaces (ABI), a smart contract interacts with users to offer the predefined business logic or contract agreement. The blockchain and smart contract enabled security mechanism for applications has been a hot topic and some efforts have been reported recently, for example, smart surveillance system \cite{nagothu2018microservice, nikouei2018realtime}, social security system \cite{xu2018constructing}, space situation awareness \cite{xu2018exploration}, identification authentication \cite{hammi2018bubbles} and access control \cite{xu2018blendcac, xu2018smartcac}. Blockchain and smart contract together are promising to provide a solution to enable a secured data sharing and access authorization in decentralized public safety systems.

\section{LISPS Framework Overview}
\label{sect:overview}


The proposed Lightweight IoT Smart Public Safety (LISPS) framework follows the divide-and-conquer principle to functionally decouple the processes in public safety and system security. The computationally expensive processes are divided into multiple sub-tasks. Based on the microservices architecture, the LISPS system offers a completely decentralized solution where sub-tasks of a function are hosted by different hardware devices. An update or change of one service does not affect the operation of the entire system as long as it follows the same input and output relations.

A Docker container is adopted in the LISPS design. The LISPS system extracts features for each object based on the movements and tracks the object of interest (e.g, pedestrians). Contextual features are added, which reflect the time and location of the camera. Based on these features, a data fusion model makes fuzzified decisions and indicates the likeliness that the objects of interest are loitering or wondering around in the frame. According to law enforcement experiences from campus police officers \cite{nikouei2018isafe}, an abnormal behavior is a good indicator pointing to some potentially risky behaviors. Timely detection of the behavior is critical to enable proactively deterrence.  

The edge processors currently available on the commercial market are not powerful enough with minimal electricity usage. The Single Board Computers (SBC) usually only contain the basic components such as the central processing unit (CPU) and memory for a low cost. SBC can handle a minimum operating system such as the kernel of Unix operating system. Using the SBCs as the edge processors, video processing and decision making are very process intensive. While our earlier tests show that the whole process cannot be done at one edge unit \cite{nikouei2018kerman, nikouei2018intelligent}, an edge device is the most ideal location to finish decision making as no delay or communication overhead is incurred. 

In the LISPS, the video processing is accomplished at the edge. No video is transferred through the network for detection of behavior, but instead, the features set is sent. Meanwhile, an Internet server is set to listen to requests for the live video, in case a human agent wants to check the real-time footage. 
Once the decision is made in the fog node, the results are shared with the authorities in the cloud who monitor the performance of the system. When the detected behavior is highly suspicious, an alarm is sent in form of an email or text massage. The LISPS design waives the need of a universal server or an app for alert sending.  

The LISPS framework includes a blockchain network for security. 
The blockchain security network serves two purposes. The devices that request and receive access to the video are preapproved and each frame's information should be protected against malicious attacks. Without relying on a centralized point of access, the blockchain network is the best candidate for distributed environments. 
In LISPS, there are trusted nodes that have access to other nodes. Based on a smart contract, these trusted nodes are connected in an access-control blockchain that is synchronized with the rest of network nodes. The second role of the blockchain network is to protect the integrity of the information shared in the network. By putting the frame information in another smart contract and sharing the hash value with the rest of the nodes, potential malicious users cannot change the feature data. 

\begin{figure}[H]
    \centering
        \includegraphics[width=0.87\textwidth]{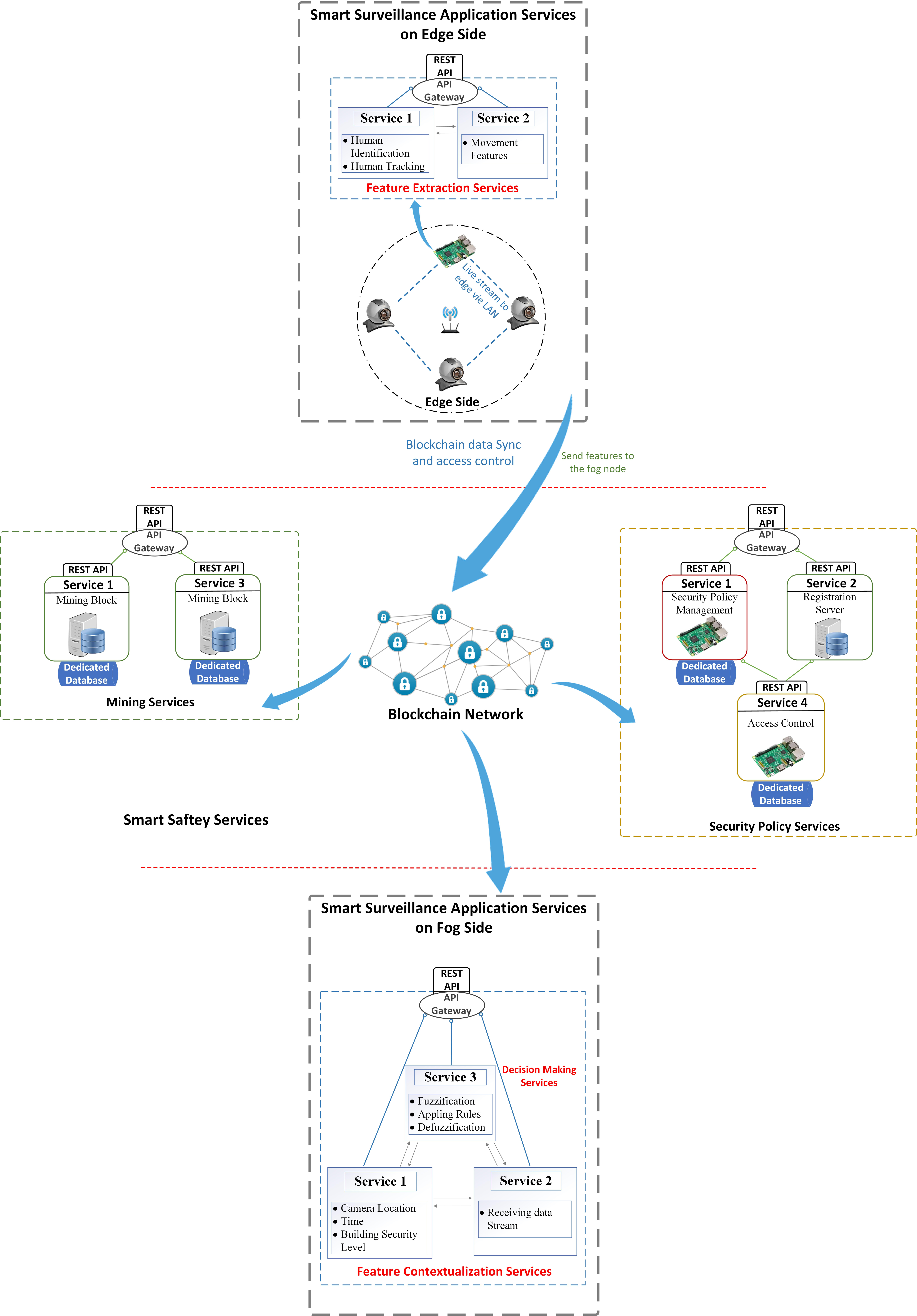}
    \caption{Illustration of the LISPS system architecture: functions deployed at the edge layer, fog layer and the security services.}
    \label{fig:sysarctecture}
    \vspace{-10pt}
\end{figure}

Figure \ref{fig:sysarctecture} illustrates the LISPS system architecture. It utilizes Docker to make the distributed connection between services while having microservices-enabled blockchain network to secure video stream services. The proposed system consists of four services:

\noindent \emph{1. Smart Safety Application Services on the Edge}: These services provide functions to support smart surveillance, such as video stream processing, object detection and tracking, and movement features extraction. Real-time video stream is generated by each camera and transmitted to a microservice running on an edge unit for features extraction. The key function of features extraction task can be decoupled into multiple microservices that are deployed on a single or multiple edge devices and work cooperatively. Lower level features are transferred to fog nodes for data aggregation and higher level analytic service. The data transfer is done using a Hyper Text Transfer Protocol (HTTP) server deployed at the host and acts as the connection between the edge and fog units.

\noindent \emph{2. Smart Safety Application Services on the Fog}: These services support the higher-level feature contextualization and smart decision making for a surveillance system. Applied on several fog nodes, these decisions are all available to the operator who should receive the alarm in case of danger. The features are saved as references on the local database of the fog node for pattern analysis. In addition, a microservice monitors the decisions. When a suspicion is identified, an email and/or text message along with the information of the alarm, are sent to the operator.
   
\noindent \emph{3. A Private Blockchain Network}: The security microservice protocol not only provides network communication channel on the Internet, but also a private blockchain network running on a decentralized peer-to-peer network. All the network communications among microservice entities that run on the TCP/IP protocol are screened for access control. Security solutions are developed as Decentralized Application (DApp), which is based on a smart contract and deployed on a blockchain network.
   
\noindent \emph{4. Blockchain-enabled Security Services}: The security service section acts as a fundamental service pool. It could be divided into two main clusters: security policy services and mining services. Each containerized miner is responsible for executing consensus algorithms to verify transactions and generating new blocks. Multiple certificated miners cooperate with each other to secure the private permissioned blockchain network. All the security polices and models, such as identity authentication and access control, are transcoded as separate microservices. Those microservices work together as a security policy services cluster to offer secured data sharing and access control services for public safety system.   

\section{Decision Making Process}
\label{sec:decison}

The LISPS platform operation consists of two phases: edge layer and fog layer processes. In this section, the details of the video feature extraction at the edge and the decision making method at the fog are presented. At the edge, a Lightweight Convolutional Neural Network (L-CNN) algorithm \cite{nikouei2018lcnn} is adopted for online object detection and the Kerman algorithm \cite{nikouei2018kerman} is used for real-time object tracking. At the fog layer a fuzzy logic based decision making scheme is proposed to detect suspicious behaviors \cite{nikouei2018isafe}. The fuzzy control system weights were determined from a representative security officer in their daily law enforcement practice. 

\subsection{The Edge Layer}

\begin{figure}[H]
    \centering
        \includegraphics[width=0.98\textwidth]{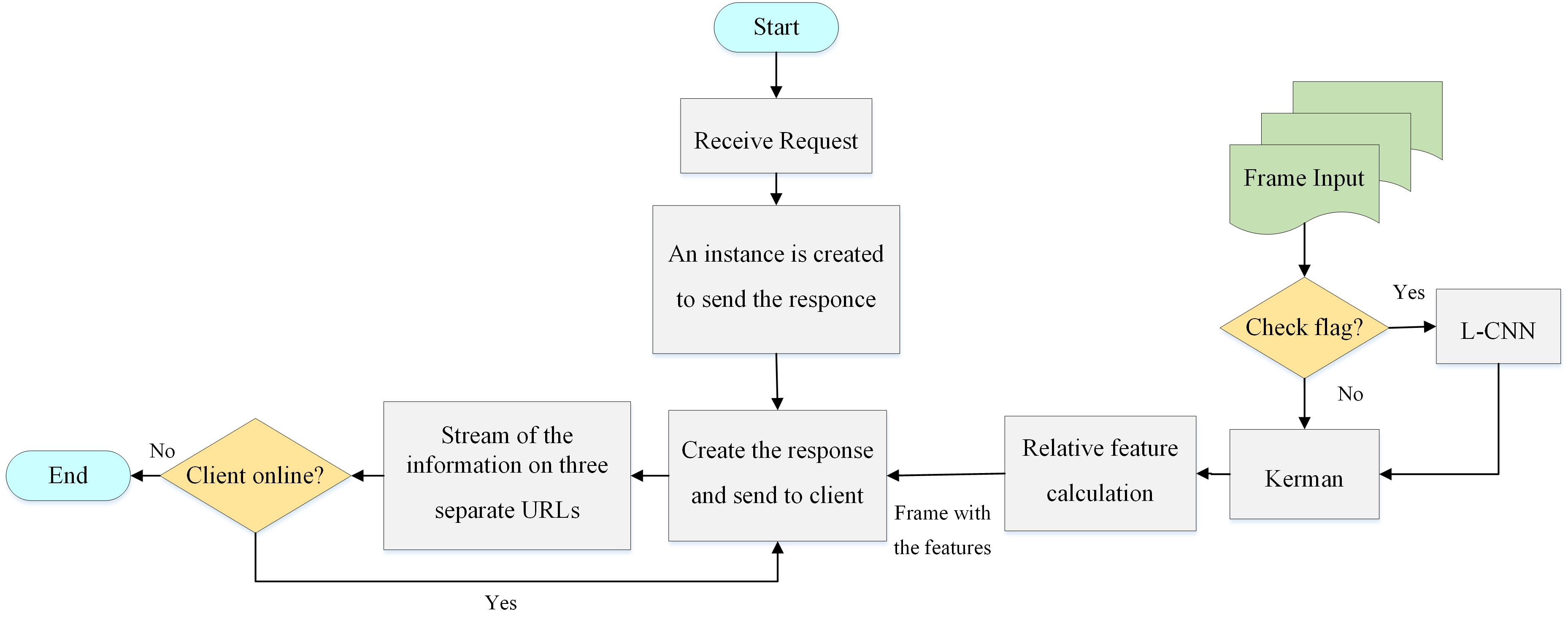}
    \caption{Edge layer video frame processing using the L-CNN and Kerman algorithms in the micro-services.}
    \label{fig:edge side}
\end{figure}

Figure \ref{fig:edge side} shows the work flow of frame processing at the edge. The video frames are captured by the camera, 
which is reachable by the microservices that read the footage. 
Multiple frames are checked every second for pedestrian detection and feature extraction. The Kerman tracker is optimized to have high accuracy in a restricted environment. If the L-CNN detects any human object in a frame, a bounding box coordination around the newly detected person is send to the queue of the tracker. If the L-CNN does not detect the pedestrian with some threshold of the Intersection over Union (IoU), form the bounding box that is already in the tracker’s queue, it is deleted from the queue and the object is considered as lost. Movement based features, such as the speed and direction changes, are extracted from the bounding box that the tracker gives for each object in each frame. 

The features for each object of interest are put into a dictionary format where the key is the object first detection time and the value consists of all the features. The dictionary is converted to a string using the pre-built libraries of python and is packed with other dictionaries for each frame to be transferred to the fog layer using HTTP. The web server handling the stream of the information is set in a multi-threaded mode where several requests for the same information are handled with the same information set and so there is no need to process the frame before stream for each request in the network. The system works on object instances that are processed in the class and share the results of the frame process.

Also, a dictionary service supports several URLs corresponding to real-time surveillance video before processing. Another microservice is set to send the visible video after processing for the human operator who wants to monitor the results of the video processing. The current feature set is transferred in text format. 


The web server on the edge side listens for incoming requests. In case the connection is lost and there are no acknowledgments, the stream is stopped and the communication can be started again with the request from the fog node. The fog-start method may save the unnecessary streaming on the network in case the receiver is gone.


\subsection{The Fog Layer}

On the fog side, the first challenge is to gather the data stream. The Python module \textit{URLLIB} (Uniform Resource Locators Library) helps to reach to the server in the edge and request for the feature data. Once the features are transmitted, the format is going to be in plain text and the fog side should wait until all the parameters in the frame are received until assessing the correct context before sending to the fuzzy mode. The receiver keeps listening until all the information belonging to a frame is delivered. The feature file is stored for a longer time in case of a need for reference. Every day, a new file is opened at time 00:00 and every ten bytes received is saved in the file. 

After the features of each object in each frame are collected, the text is transformed back to the features in the format of the dictionary. Before a decision is made, the data should be contextualized. Several factors can be considered for contextualization, three most common ones are: the time in the day, the location of the camera, and the security level of the building that the camera is located in. 
Similar to the architecture that was explained in the edge layer, the rest of the functions on fog level are implemented in Docker containers for ease of distribution or scalability. The Docker image may have all the components and dependencies that are needed for the code to run. Each byte of a string is one string character.

\begin{figure}[H]
    \centering
        \includegraphics[width=0.98\textwidth]{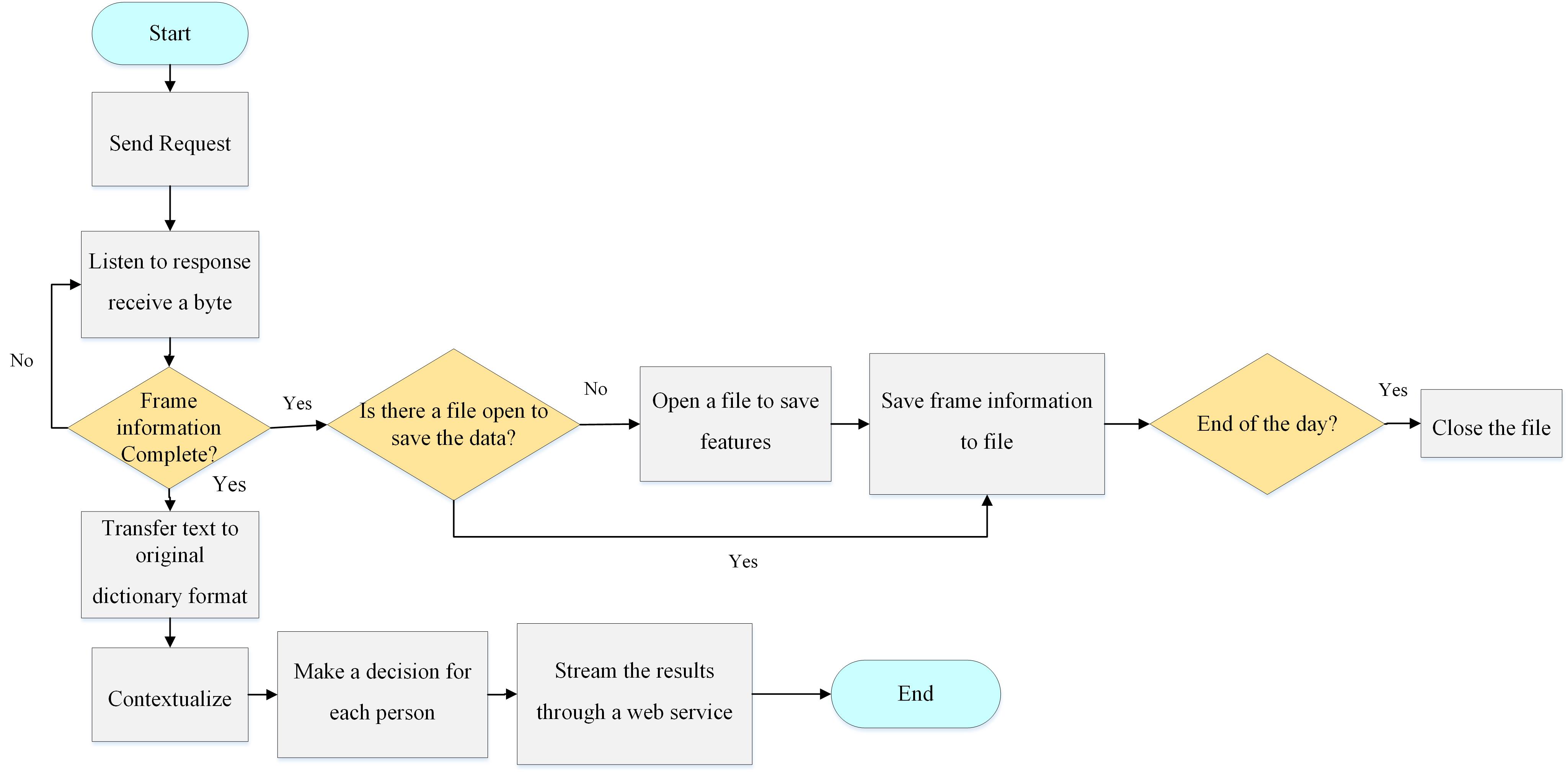}
    \caption{Fog layer decision making working flow in the microservices.}
    \label{fig:fog side}
\end{figure}

Figure \ref{fig:fog side} shows the steps that the fog node takes to make the decision. After the contextualization, the features of each detected human object in the frame are divided and separately given to a fuzzy control system. It returns the suspicious activity level regarding a person anomaly after defuzzification. A threshold is defined by the system administrator based on the past experiences. When the suspicion level of an individual object is below the pre-set threshold, the results are ignored. Otherwise, an email or a text massage is sent to the operator. The LISPS framework is a decentralized system, with an email service sends emails or text massages only to designated receiver, instead of flooding to the entire system. The email or text message contains the pre-recorded alarm along with the features, time and camera ID, helping the operator finds the scene easily. 

\subsection{Secured Data Access Control}

All the entities on the permissioned blockchain network are implemented as containers, which perform blockchain services independently on the host devices. The containerized microservices could be categorized as miners or non-mining nodes given the computation power of the host devices. An identity management mechanism ensures a new node enrollment process in the permissioned blockchain network. Only authorized participants could be recognized by entities of the network and perform blockchain services, such as mining blocks, sending transactions and deploying smart contracts. Compared to a public blockchain network, the \emph{permissioned blockchain network} achieves higher efficiency in consensus operation, and more secured network by limiting participants and clear defined security policies. 

Utilizing the microservices architecture, the security functions are decoupled into multiple microservices and deployed on distributed computing devices. These decentralized security microservices work as a service cluster to offer a scalable, flexible and lightweight data sharing and access control mechanism for the LISPS system \cite{xu2019blendmas}. An entity registration process is performed by the registration microservices associating entity's unique blockchain account address with a Virtual ID (VID). The identity authentication microservices expose RESTful APIs to other microservices-enabled providers for referring identity verification results. 

The security management microservices act as data and security service managers who deploy the smart contracts encapsulating hashed index authentication (HIA) and the access control policies. In the hashed index recording process, the hashed index recording (HIR) microservices simply interact with the authorized ABI function to save the hashed index data to the blockchain. When it comes to hashed index authentication scenario, microservice queries a hashed key-value index by interacting with smart contract and compares it with calculated hash values of the record index table, the return the verification result. The access control microservices encapsulate access control model and perform access right validation during service granting process on smart surveillance service providers.

\section{Experimental Study}
\label{sec:exp}

\subsection{Experimental Setup}

The docker image is deployed on two types of SBCs, a TinkerBoard and a Raspberry PI 3 model B+. The Tinker Board has 1.8 GHz ARM-based RK3288 SoC and 2 GB LPDDR3 dual-channel. The Raspberry PI 3 carries a 1.4 GHz CPU, which is a Cortex-A53 (ARMv8) 64-bit SoC, and a 1 GB of LPDDR2 memory. Both have an ARM based CPU that makes it easier to deploy a Docker image based on the CPU architecture on both. The characteristic of low power consumption along with the small size and portability makes the SBCs the major candidates for edge processing. 

A laptop is adopted as the fog node, which has a 8-th generation Intel core i7 processor @3.1GHz with 4 physical cores and 16 GB of DDR4 memory. Even a cellphone can be considered a fog node, and in this scenario, a laptop is considered for testing as a middle point between a personal computer (PC) and a mobile phone processor. The design takes advantage of the multi-threads offered by the laptop.

\subsection{Experimental Results}
\begin{figure}[ht]
    \centering
        \includegraphics[width=0.68\textwidth]{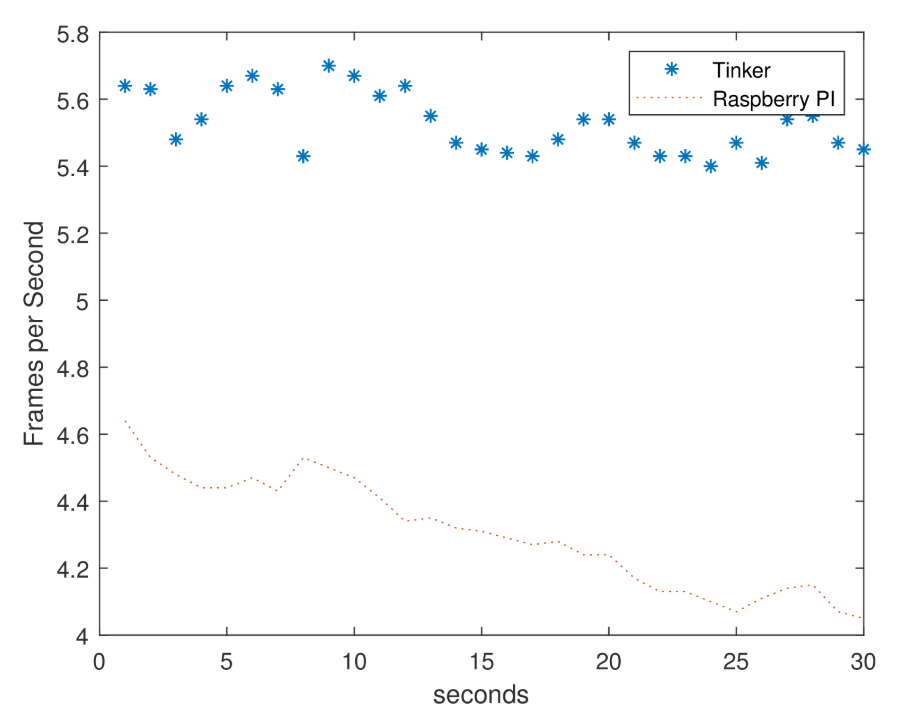}
    \caption{The speed of feature extraction at the edge of the network in Raspberry PI and Tinker Board.}
    \label{fig:fps}
\end{figure}

Figure \ref{fig:fps} compares the performance of the edge units, implemented on Raspberry PI and Tinker board respectively, in terms of Frames Per Second (FPS). The edge nodes processed video streams of 30 seconds and the rates are recorded. Compared to the performance of the same algorithm but implemented on the host outside of the Docker container \cite{nikouei2018kerman}, an average decrease of 1.9 FPS is observed for the Tinkerboard. The trade-off paid for the modularity offered by the microservices. An average of 5.5 FPS achieved on these edge units, which is sufficient for “real-time” public safety applications, especially taking the velocity of pedestrians into account. Meanwhile, the average FPS of Raspberry PI is around 4 FPS.. Better performance can be expected as new hardware is utilized in the syste design. For example, a Raspberry PI model B+ board, offer a faster and more reliable processor. 

The container for video frame processing takes approximately 246 MB memory when deployed, although the image of the container can be more than 5 GB. The measurement shows the feasibility to migrate the video frame processing function to the edge of the network. For example, even a Raspberry PI board with 1 GB of memory is able to handle the container image and still has sufficient memory remained for buffering and a blockchain container. The Raspberrian operating system of the Raspberry PI is a down-graded Debian, which does not require much memory to operate while performing vital tasks such as the power management and down clocking the CPU.

\begin{figure}[ht]
    \centering
        \includegraphics[width=0.68\textwidth]{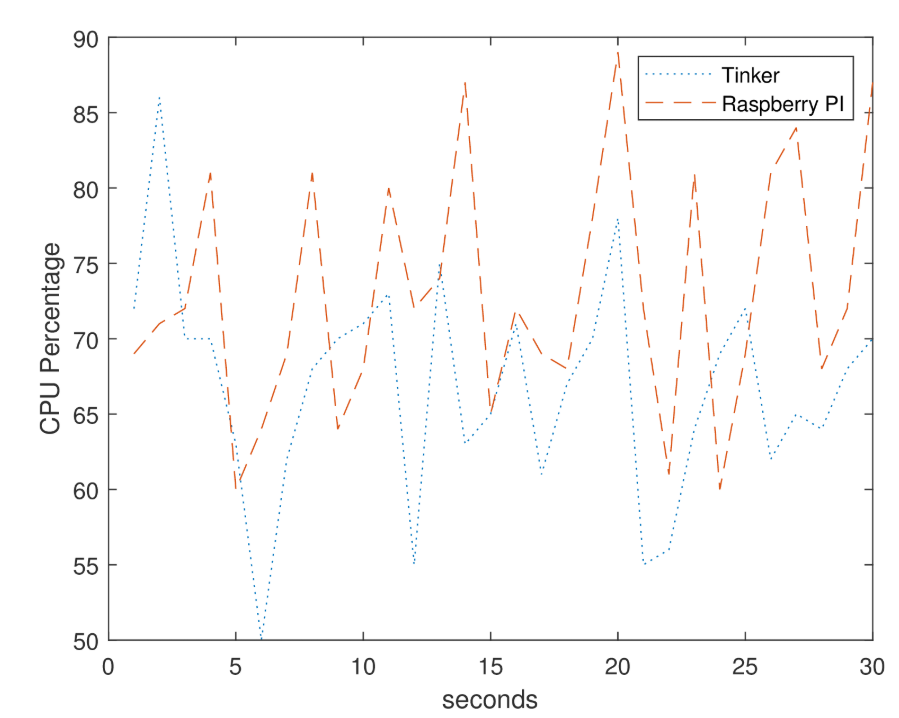}
    \caption{CPU usage percentage in the edge units.}
    \label{fig:cpu}
\end{figure}

CPU utilization is another important parameter as the edge unit works with a much lower power and slower clock rate using the ARM (Advanced RISC (reduced instruction set computer) Machines) architecture. The CPU manages to be less than 80\% in average. It is high, but maintainable for the device. It appears that the same algorithm has access to more resources when run outside of the container, but the Docker enforces some limitations on how much resource is available to each container.  Figure \ref{fig:cpu} shows the CPU usage recorded in 30 seconds of run time of the video processing container in each of the edge devices with no other processes running. 

\begin{figure}[t]
    \centering
        \includegraphics[width=0.68\textwidth]{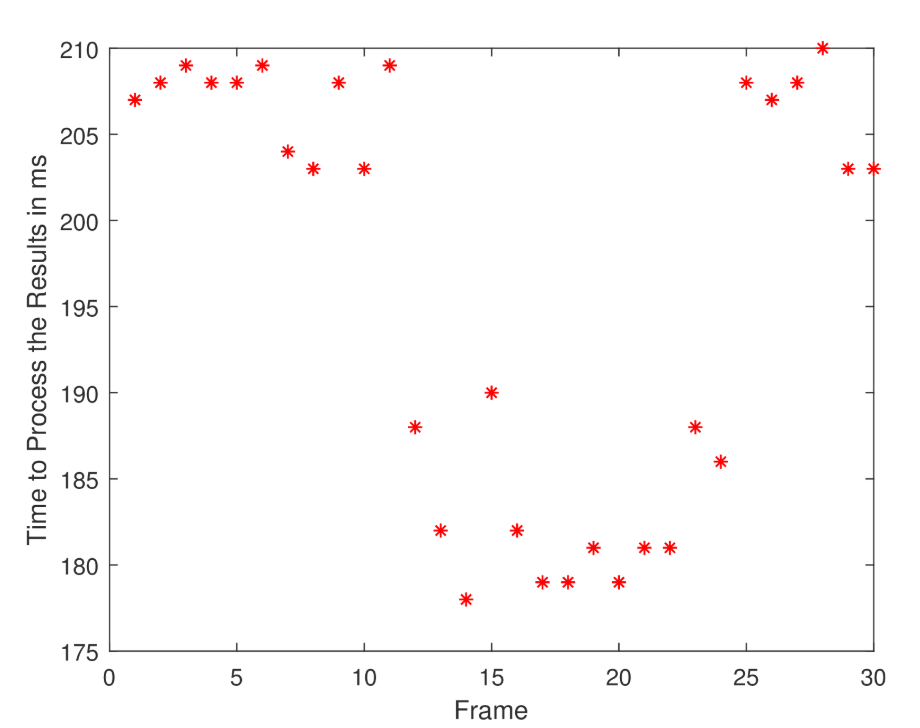}
    \caption{Time taken to process each of the received frames on the fog node, including the idle time that the fog node takes to wait until the data of the frame is received completely.}
    \label{fig:fognode}
    \vspace{-10pt}
\end{figure}

On the fog unit, the container is handled much easier on the laptop hardware than expected. The memory consumption by the algorithm is not as high as the video processing model on the edge as there is not a deep learning architecture. The container itself may be smaller in size too because the libraries for video processing are not needed anymore. The fog node uses one of the CPU cores closer to 90\% to handle the video stream received by a single edge node. 
Figure \ref{fig:fognode} shows the time for processing of 30 frames in milliseconds at the fog in real-time. However, in a sense, the interval also incorporates the time for edge node to extract and send features to the fog unit before fog can process the features. 

A speed degradation is observed when there are as many containers running as CPU threads on the fog node. It is assumed that the background processes are pushed to a utilized CPU thread that should otherwise be used for decision making, resulting in the slowdown. Thus, it is recommended to have \emph{fewer containers deployed than the number of CPU threads}. It should be mentioned that having more containers running does not result in the system crash. 

\section{Conclusions and Discussions}
\label{sec:conclusions}

The LISPS system is motivated to provide a lightweight decentralized platform for video processing for smart public safety with minimum delay and network dependencies and easy to deploy and configure. To cope with the quick-changing, heterogeneous and dynamic IoT infrastructure, continuous updates and integration of new features without interrupting system operation is desired. Leveraging the microservices architecture, the LISPS framework delivers an uninterrupted experience in which complex algorithms are broken down to smaller and simpler services. 

Based on our previously proposed lightweight object detection, tracking, and suspicious behavior detection algorithms, the human-oriented public safety system architecture is taken as a case study and implemented using Docker containers. On the edge device, two containers are deployed, one is based on smart contracts in a private blockchain network that ensures the data integrity in communication, and the other container is in charge of \emph{surveillance video processing}.

Decision making for several edge nodes in the same geo-location is outsourced to a fog unit. The fog node runs multiple containers and each of them is responsible for different sub-tasks, from collecting the edge device stream to decision-making based on the features, along with consensus algorithm and smart contract management required to enforces blockchain protocol. 

The experimental achievements verifies that the LISPS prototype takes $< 0.5$ second to detect a suspicious person in a video frame and send a message to the authorities. Considering the velocity of pedestrians, even this low resolution is acceptable as a “real-time” processing. In case there are large number of people in the scene, the processing and decision making may take longer time. The LISPS prototype presented here with the described hardware is capable of maintaining the performance with up to five objects in the frame at the same time, beyond which the delay may be noticeable. 


There are several areas for further improvements. More features are needed to have a better detection from which other techniques can be considered \cite{liang2015encoding}. Another part includes an encrypted channel that ensures data confidentiality by preventing unauthorized third parties from accessing the raw video or feature data in transmission. 




\bibliography{report}   
\bibliographystyle{spiejour}   

\end{document}